\documentclass{article}
\usepackage{graphicx}  
\usepackage{amsmath}   
\usepackage[compress]{cite}
\usepackage{amssymb}   
\usepackage{bm} 
\usepackage{dcolumn}
\usepackage{color}
\usepackage{mathrsfs}
\usepackage{amsfonts}
\usepackage{enumerate}
\usepackage{varioref}
\RequirePackage[colorlinks,citecolor=blue,urlcolor=magenta,linkcolor=blue]{hyperref}

\def\LL{Lanczos-Lovelock }
\def\gr{general relativity}
\newcommand{\nn}{\nonumber}
\newcommand{\vbar}{\left\vert\vphantom{\int}\right.}
\labelformat{section}{Section #1} 
\labelformat{subsection}{Section #1} 
\labelformat{subsubsection}{Section #1}
\labelformat{subsubsubsection}{Section #1}
\labelformat{equation}{Eq.~(#1)} 
\labelformat{figure}{Fig.~#1} 
\labelformat{subfigure}{Fig.~\thefigure#1} 
\labelformat{table}{Tab.~#1} 
\labelformat{appendix}{Appendix #1}
\title{Gravitational field equations near an arbitrary null surface expressed as a thermodynamic identity}
\author{ 
		{\bf {\normalsize Sumanta Chakraborty$^a$,}$
			$\thanks{E-mail: sumanta@iucaa.in, sumantac.physics@gmail.com}} \, 
		{\bf {\normalsize Krishnamohan Parattu$^a$,}$
			$\thanks{E-mail: krishna@iucaa.ernet.in}} \,\\\\ and
		{\bf {\normalsize T. Padmanabhan$^a$}$
			$\thanks{E-mail: paddy@iucaa.in}}\\\\
		{\normalsize $^a$IUCAA, Post Bag 4, Ganeshkhind,}
		\\{\normalsize Pune University Campus, Pune 411 007, India}
		\\[0.3cm]
	}
\begin{document}
  
\maketitle
\begin{abstract}
Previous work has demonstrated that the gravitational field equations in all \LL\ models  imply a thermodynamic identity $T\delta_\lambda S=\delta_\lambda E+P\delta_\lambda V$ (where the  variations are interpreted as changes due to virtual displacement along the affine parameter $\lambda$) in the near-horizon limit in static spacetimes. 
Here  we generalize this result to \textit{any  arbitrary null surface in an arbitrary spacetime} and show that certain components of the Einstein's equations can be expressed in the form of the above thermodynamic identity. We also obtain an explicit  expression for the thermodynamic energy associated with the null surface. Under appropriate limits, our expressions reduce to those previously derived in the literature.
The components of the field equations used in obtaining the current result are orthogonal to the components used previously to obtain another related result, viz. that some components  of the field equations reduce to a Navier-Stokes equation on any null surface, in any spacetime. We also describe the structure of Einstein's equations near a null surface in terms of three well-defined projections and show how the different results  complement each other. 
 
\end{abstract}
\section{Introduction}\label{Paper04:Sec:01}

Horizons in general (and black holes in particular) possess thermodynamic attributes like entropy \cite{Bekenstein:1973ur,Bekenstein:1974ax} and temperature \cite{Hawking:1974sw,Davies:1976ei,Unruh:1976db,Gibbons:1976ue}. These features, which are known to transcend Einstein's gravity, are believed to stem from some deep connection between gravitational dynamics and horizon thermodynamics. In recent years, it has been shown that in \gr\ \textit{as well as in a wider class of gravitational theories},  the field equations near a horizon imply a thermodynamic identity $T\delta_\lambda S=\delta_\lambda E+P\delta_\lambda V$ \cite{Padmanabhan:2002sha,Padmanabhan:2003gd,Padmanabhan:2009vy,Cai:2005ra,Akbar:2006er,Kothawala:2007em,Paranjape:2006ca,Cai2008,Kothawala:2009kc,Akbar2009,Akbar2007,Cai2008a,Cai2007,Akbar2007a,Cai2007a,Gong2007,Wu2007,Cai2009,Cai2010,gravitation,Wald:1999vt,Padmanabhan:2007xy,Padmanabhan:2002xm,Padmanabhan:2002jr} where the symbols have the usual meanings and the variations are interpreted 
as changes due to virtual displacement along the affine parameter $\lambda$. 
This result --- originally obtained for \gr \cite{Padmanabhan:2002sha} --- has been generalized to all static spacetimes with horizon in  the \LL theories of gravity \cite{Padmanabhan:2013xyr,Paranjape:2006ca,Kothawala:2009kc}. 
That is, the result is known to hold for actual horizons --- rather than to generic null surfaces --- and requires the assumption of a static spacetime.

It may therefore appear that this connection --- between the field equations and a thermodynamic identity --- is an exotic phenomenon that occurs only in specific solutions containing  horizons. But this illusion is broken when we realize that a generic null surface through any event in spacetime can act as a local Rindler horizon for some observer \cite{Jacobson:1995ab, Padmanabhan:2009vy}. This fact allows one to introduce observer-dependent thermodynamic variables around any event in spacetime and 
reinterpret the gravitational field equations near any null surface in a thermodynamic language. One can then `derive' field equations in the case of Einstein's theory  \cite{Jacobson:1995ab} from the Clausius relation applied to a null surface, if one assumes further that (a) the entropy density is one quarter of the transverse area and, more importantly, (b) the quadratic terms in the Raychaudhuri equation --- involving the squares of shear and expansion --- can be set to zero. (This approach based on the Raychaudhuri equation, however, could not be generalized in a simple manner to more general class of theories.) A clearer connection between null surface thermodynamics and gravitational dynamics emerged from the fact that, gravitational field equations reduce to the Navier-Stokes equations of fluid dynamics in \textit{any} spacetime when  projected on an \textit{arbitrary} null surface \cite{Padmanabhan:2010rp,Kolekar:2011gw}, thereby generalizing previous results for black hole horizons \cite{Damour:
1982}.

In the light of these results, it is natural to ask whether gravitational dynamics is the long wavelength thermodynamic limit of the dynamics of some unknown microscopic degrees of freedom. 
The conceptual framework that attempts to interpret gravitational dynamics as emergent from the dynamics of unknown microscopic degrees of freedom is known as the emergent gravity paradigm. It has received significant amount of support from later investigations, especially from the following results: 
(i) It is possible to express the action functional for gravity as the sum of a bulk term and a surface term with a ``holographic" relation between them. This result holds not only in Einstein gravity but also in all \LL theories of gravity \cite{Mukhopadhyay:2006vu,Kolekar:2010dm,Kolekar:2011bb}.
(ii) More recently, it has been shown \cite{Padmanabhan:2013nxa} that the total Noether charge in a 3-volume $\mathcal{R}$ related to the time evolution vector field can be interpreted as the heat content of the boundary $\partial \mathcal{R}$ of the volume and the time evolution of the spacetime itself can be described in an elegant manner as being driven by the departure from holographic equipartition measured by  $(N_{\rm bulk} - N_{\rm sur})$. Here, the number of bulk degrees of freedom $N_{\rm bulk}$ is related to the Komar energy density while the number of surface degrees of freedom $N_{\rm sur}$ is related to the geometrical area of the boundary surface. All these results generalize in a non-trivial manner to \LL theories of gravity \cite{Chakraborty:2014rga}.

These facts suggest that the gravitational field equations on (or near) \textit{any} null surface in \textit{any} spacetime might have a natural thermodynamic interpretation. One should be able to understand all the previous results as different facets of a unified picture and also generalize them to an arbitrary null surface. In particular, it should be possible to obtain from Einstein's equations the thermodynamic identity $T\delta_\lambda S=\delta_\lambda E+P\delta_\lambda V $ near \textit{any} null surface, thereby freeing the earlier demonstrations from their  restrictive assumptions, like the assumption of spacetime being static \cite{Kothawala:2009kc} or having a specific horizon. In this paper, we will provide such a  generalization of these results for an arbitrary null surface in an arbitrary spacetime, which is neither static nor spherically symmetric. 

As we have mentioned earlier, it is possible to attribute thermodynamical entities like temperature and entropy to any null surface by introducing local Rindler horizons. Then, with a suitably defined pressure $P$ (or more precisely, a work function), we can consider an infinitesimal displacement of the horizon along the affine parameter $\lambda$ of the null geodesics off the surface and show that Einstein's equations imply the relation:
\begin{equation}
 T\delta _{\lambda}S-P\delta _{\lambda}V= \delta _{\lambda}E 
\label{first}                                                           
\end{equation}  
for a suitably defined energy $E$. Since we already have well-defined, physically motivated, expressions for $T,S,P$ and $V$, it is now possible to identify the \emph{energy} associated with the null surface,
which appears in this thermodynamic identity for an arbitrary null surface under consideration. Further, starting from this result and taking suitable limits we can arrive at the previous results in the literature as special cases. 

We will obtain the result in \ref{first} using the component of the Noether current along the null geodesics on the surface and the relation between Ricci scalar for the full spacetime with the Ricci scalar for the two-surface. This provides a physically well motivated route to study thermodynamic structure of the spacetime which has proved to be quiet useful in the past \cite{Padmanabhan:2013nxa,Chakraborty:2014rga}.

The paper is organized as follows: In \ref{Paper04:Sec:02}, we will introduce the Gaussian null coordinates near an arbitrary null surface which we will use throughout the paper. Then, in \ref{sec:GRtherm1}, we describe three natural projections of Einstein's equation which arise in the presence of a null surface and their physical interpretation. In \ref{sec:GRtherm}, we introduce the Noether current associated with a vector field from a simple geometric identity (which does not require talking about diffeomorphism invariance of an action etc.) and identify the vector for which we shall compute the Noether current. Next we evaluate the component of the Noether current along the null geodesics on the null surface and use it to derive the thermodynamic identity by considering a virtual displacement of the null surface along the auxiliary null geodesics. Subsequently, in \ref{sec:special}, we reproduce the previous results  available in the literature by specializing to stationary and static metrics as well as 
to 
spherically symmetric (but not necessarily static) metrics.  Finally, we conclude with a short discussion of our results.

Throughout this paper, we use metric signature $\left(-,+,+,+\right)$ with the fundamental constants $G$, $\hbar$ and $c$ being set to unity so that Einstein's equations reduce to  $G_{ab}=8\pi T_{ab}$. The Latin indices a,b,$\ldots$ run from $0$ to $3$ and stand for spacetime coordinates, Greek indices $\mu$, $\nu$, $\ldots$ run from $1$ to $3$ and represent coordinates on the null surface and capitalized Latin indices A,B,$\ldots$ stand for coordinates on the $2$-surface transverse to the normal to the null surface and its auxiliary null vector.

\section{Gaussian Null Coordinates (GNC)}\label{Paper04:Sec:02}

Since we are interested in the form of the field equations near a null surface, we will begin by introducing a coordinate system $(u,r,x^A)$ adapted to the null surface. This coordinate system will be constructed in close analogy with what we expect in a local Rindler frame and will have the following properties: (a) There will be only 6 free functions in the metric thereby freezing all redundant gauge degrees of freedom. (b) The null surface we are interested in is chosen to be a surface $r=0$. Further, $r=$ constant but non-zero  surfaces will represent timelike surfaces with $r\to0$ leading to the null surface we are interested in as a limit. (c) Observers at rest in this spacetime with constant values for $(r,x^A)$ will be analogous to local Rindler observers and will perceive $r=0$ as their local Rindler horizon. Let us briefly review how such a coordinate system can be constructed around any null surface in any spacetime.

Any arbitrary null surface can be parametrized using Gaussian null coordinates (henceforth referred to as GNC), which can be constructed in analogy with standard Gaussian normal coordinates associated with, say, a spacelike surface. In the \textit{non-null} case, the construction proceeds by using geodesics normal to the surface. This construction breaks down in the null case, since  geodesics with tangent vectors along the surface normal, are actually on the null surface. This problem is avoided by introducing an auxiliary null vector $k^{a}$, satisfying $\ell _{a}k^{a}=-1$ where $\ell _{a}$ is normal to the surface, and then constructing the coordinates by moving away from the null surface along the null geodesics of $k^{a}$. The construction of this coordinate system has been detailed in \cite{Moncrief:1983,Morales,Parattu:2015gga} and we will only recall its essential properties. The line element adapted to an arbitrary null surface (identified with $r=0$) takes the following form in GNC:
\begin{equation}\label{Paper04:Sec:02:Eq.01}
ds^{2}=-2r\alpha du^{2}+2dudr-2r\beta _{A}dudx^{A}+q_{AB}dx^{A}dx^{B}
\end{equation}
This line element contains six independent parameters $\alpha$, $\beta _{A}$ and $q_{AB}$, all dependent on all the coordinates $\left(u,r,x^{A}\right)$.  The metric on the two-surface (i.e. $u=\textrm{constant}$ and $r=\textrm{constant}$) is represented by $q_{AB}$. The surface $r=0$ is the fiducial null surface but surfaces with $r=$ non-zero constant are not null. The  normal $\ell _{a}= \partial_a r$ to the $r=$ constant surfaces will be a null vector on the $r=0$ null surface. Hence, the null normal  $\ell ^{a}$ and the corresponding auxiliary null vector $k^{a}$ have the following components \cite{Parattu:2015gga} in this coordinate system:
\begin{subequations}
\begin{align}
\ell _{a}&=\left(0,1,0,0\right),\qquad \ell ^{a}=\left(1,2r\alpha +r^{2}\beta ^{2},r\beta ^{A}\right)
\label{Paper04:Sec:02:Eq.02a}
\\
k_{a}&=\left(-1,0,0,0\right),\qquad k^{a}=\left(0,-1,0,0\right)
\label{Paper04:Sec:02:Eq.02b}
\end{align}
\end{subequations}
While dealing with an arbitrary null surface with normal $\ell_a$, there is some freedom in the choice of $k^a$. In our case, we have chosen the auxiliary vector as the vector that was used in the construction of the GNC \cite{Parattu:2015gga} itself, and hence, up to a sign, is a basis vector in the GNC. Once the coordinate system adapted to the the null surface is fixed, this particular $k^a$ is specified by the conditions $k_a e^a_{A}=0$  in addition to the usual conditions on an auxiliary vector, $k^a k_a=0$ and $k^a \ell_a=-1$. (Here, $e^a_{A}$ (with $A=1,2$) denote the two coordinate basis vectors on the $u=\textrm{constant}$ $2$-surfaces on the null surface.) This allows us to work with a physically well-defined basis $\{\ell_a,k_a, e^a_{A}\}$. Later on, we will take the projection of certain vectors along these basis vectors. The choice made here allows us to take the projections which have direct thermodynamic meaning; if we use a linear combination of these vectors or make some other choices, then, 
of course, the projections will get mixed up with each other and one cannot provide a simple interpretation to them.

The non-affinity parameter $\kappa$ for the null normal $\ell ^{a}$ is defined via the relation $\ell ^{b}\nabla _{b}\ell ^{a}=\kappa \ell ^{a}$. It turns out that the non-affinity parameter for the null normal we are considering is $\kappa =\alpha$ thereby allowing us to interpret the $\alpha$, which occurs in the metric in \ref{first}, as the surface gravity. The vector $k^{a}=-\partial /\partial r$ is tangent to the ingoing null geodesic (ingoing since it points in the direction of decreasing $r$), which is affinely parametrized with affine parameter $r$. We denote $\lambda _{H}$ to be the value of the affine parameter on the null surface. In the remaining discussions, we will work with $\lambda$ defined through the following relation: $r=\lambda -\lambda_{H}$.
It is also useful to introduce the vector field:
\begin{equation} \label{xi}
\boldsymbol{\xi} = \frac{\partial}{\partial u} = \left(1,0,0,0\right)~.
\end{equation}
which goes to $\ell^a$ on the null surface. This vector is special since it corresponds to the standard time direction in some well-known spacetimes that can be obtained as special cases of GNC (see \ref{app:static}) and in the local Rindler frame. We shall  describe the dynamics of the null surface from the point of view of observers moving along the integral curves of $\xi^a$ in the region $r>0$. (We have arbitrarily chosen one side of the null surface. We could have as well chosen the $r<0$ side.) Thus, $\xi^a$, representing the time direction for our fiducial observers, has to be timelike in the $r>0$ region. In order to achieve this property for $\xi^a$, we shall assume that
$
\alpha > 0,
$
at least near the null surface in the $r>0$ region. (This restriction is also consistent with the idea of identifying  $\alpha \vert_{r=0}$ to be the surface gravity and associating a temperature $\alpha/2\pi$ with the null surface.)
\section{Projections of Einstein's equations}\label{sec:GRtherm1}

The vector $\xi ^{a}$, introduced earlier, when normalized, gives the four-velocity $\xi^a/|\xi|$ a fundamental  observer with $r,\theta ,\phi =\textrm{constant}$ in the spacetime described by the metric in \ref{first}. Further, on the null surface, $\xi ^{a}\rightarrow \ell ^{a}$. Therefore the flux of the matter energy momentum tensor through the null surface is determined by the four momentum:
\begin{align}
S^{a}\equiv T^{a}_{b}\xi ^{b}\rightarrow T^{a}_{b}\ell ^{b}=T^{a}_{u}~.
\end{align}
where the second relation holds in the limit of $r\to0$. When field equations $G_{ab}=8\pi T_{ab}$ hold, we find that
$S^{a}=T^{a}_{b}\ell^{b}=(1/8\pi)G_{ab}\ell^{b}$ on the null surface. Algebraically, however, it turns out to be simpler to concentrate on the Ricci tensor rather than Einstein tensor and define a closely related vector:
\begin{equation}
 P^{a}\equiv 2R^{a}_{b}\ell^{b}=16\pi \bar {T}^{a}_{b}\ell^{b}; \qquad \bar T^a_b\equiv T^a_b-\frac{1}{2}\delta^a_bT
\end{equation} 
The structure of Einstein's equation near a null surface is determined by the components of $P^a$ on the null surface.
To investigate these components, let us expand $P^a$ in the orthonormal basis \cite{Parattu:2015gga} made of $(\ell^a,k^a,e^A)$ as $P^a=\phi_1\ell^a+\phi_2k^a+\phi_Ae^A$. This allows us to
construct two scalars $(\phi_1,\phi_2)$ and one transverse vector $\phi_A$ from the combinations: $\phi_1=-P^ak_a,\phi_2=-P^a\ell_a$ and the projection $P^{a}q_{a}^{b}$.
(Of these, $P^{a}k_{a}$ and $P^{a}q_{a}^{b}$ together represent the three components of the projection of the flux on the null surface while $P^a\ell_a$ brings out the component along $k^a$ which is the tangent vector to the \textit{ingoing} null geodesic.) 

\textit{Remarkably enough, these three components $(P^{a}q_{a}^{b}, P^a\ell_a, P^a k_a)$ lead to the three sets of results obtained in the literature earlier.} The first one ($P^{a}q_{a}^{b}$) leads to the Navier-Stokes equation, the second ($P^a\ell_a$) is related to Raychaudhuri equation and the associated results while the third one $(P^ak_a)$ leads to the thermodynamic identity in the special cases considered earlier in the literature. We will briefly describe these three and then investigate the last one in detail.

\begin{itemize}

\item \textit{The contraction $P^a\ell_a$:} Contraction of the momentum $P^{a}$ with the null generator of the null surface, $\ell ^{a}$, leads to the standard Raychaudhuri equation, 
\begin{equation}
R_{ab}\ell ^{a}\ell ^{b}=-\frac{1}{2}\Theta ^{2}-\sigma _{ab}\sigma ^{ab}+\omega _{ab}\omega ^{ab}-\dfrac{d\Theta}{d\lambda},
\end{equation} 
involving the combination $R_{ab}\ell ^{a}\ell ^{b}$.
It is this Raychaudhuri equation which was used in the work by Jacobson \cite{Jacobson:1995ab}, along  with the crucial assumption of vanishing $\Theta$ and $\sigma$ for the chosen null surface, to obtain Einstein's equation from Clausius relation, i.e., $\delta Q=TdS$. In other words, the component of $P^a$ responsible for Jacobson's thermodynamic interpretation of Einstein's equations is obtained by the contraction of $P^a$ with the null generator $\ell _{a}$.
But, for providing this interpretation, one needs to make additional restrictive assumptions (like the vanishing of shear and expansion) which will \textit{not} hold on an \textit{arbitrary} null surface. Later on, some of these restrictive assumptions in Jacobson's work were lifted in \cite{Chirco2009} but this demanded the interpretation of the shear and expansion terms as dissipative effects. A more detailed discussion of the differences between Jacobson's approach and our approach, as well as some crucial issues in the former are highlighted in \cite{Kothawala:2010bf}. It should also be noted that the approach based on the Raychaudhuri equation cannot be generalized  in a simple manner to Lanczos-Lovelock models, while it turns out that our approach does generalize in a straightforward manner to \LL theories of gravity \cite{Chakraborty:2015wma}.

We want to work with an arbitrary null surface with non-zero shear and expansion and  we want to develop an approach which will generalize to \LL\ models in a natural fashion. It, therefore, turns out to be more fruitful to study the projection of $P^a$ orthogonal to $\ell_a$, especially the contraction $P^ak_{a}$.   As we shall show, the contraction along $k^a$ has a neat thermodynamic interpretation without  additional assumptions or introduction of dissipative effects.  The thermodynamic interpretation of the contraction with $P^ak_{a}$ also generalizes in a straightforward manner to \LL theories of gravity \cite{Chakraborty:2015wma}.  In short, the projection on $k_{a}$ leads to richer thermodynamic content. The physical reason for this could be the following: Note that, since $\ell^2=0$ and $\ell^ak_a=-1$, the contraction $P^ak_{a}$ actually  picks out the components of $P^a$ along $\ell^a$  which is intrinsic to the null surface.

\item \textit{The contraction $P^{a}q_{a}^{b}$:} Let us start with contraction of $P^{a}$ with the transverse metric $q_{ab}$ which is proportional to $R_{ab}\ell ^{a}q^{bc}$. This expression --- when worked out in detail --- leads to the Navier-Stokes equation on the null surface \cite{Padmanabhan:2010rp}.  More specifically, using vectors and derivatives intrinsic to the null surface, the contraction of $P^{a}$ with $q_{ab}$ leads to \cite{Padmanabhan:2010rp}
\begin{align}\label{Dissipation_2010}
R_{mn}\ell ^{m}q^{n}_{a}&=q^{m}_{a}\pounds _{\ell}\Omega _{m}+\Theta \Omega _{a}-D_{a}\left(\kappa +\frac{\Theta}{2}\right)+D_{m}\sigma ^{m}_{a}=8\pi T_{mn}\ell ^{m}q^{n}_{a}
\end{align}
where $D_{a}$ is the covariant derivative defined on the null surface using the projector $q_{ab}$ and $\pounds _{\ell}$ denotes the Lie derivative along the null generator $\ell$. We have also separated out the trace of $\Theta _{ab}$ and have defined a new object $\sigma _{mn} = \Theta _{mn}-(1/2)q_{mn}\Theta$. It is clear from \ref{Dissipation_2010} that it has the form of a Navier-Stokes equation for a fluid with the convective derivative replaced by the Lie
derivative. This correspondence allows us to give the following interpretations to geometric quantities on the null surface: (i) The momentum density is given by $-\Omega _{a}/8\pi$ where $\Omega _{a}=\kappa k_{a}+\ell ^{j}\nabla _{j}k_{a}$. In the coordinates adapted to the null surface, $\Omega_a$ has only transverse components which are given by $\Omega_A=\beta_A$; this suggests interpreting $\beta_A$ as the transverse fluid velocity. Further, we have identified the (ii) pressure $\kappa/8\pi$, (iii) shear tensor defined as $\sigma _{mn}$, (iv) shear viscosity coefficient $\eta = (1/16\pi)$, (v) bulk viscosity coefficient $\zeta = -1/16\pi$ and finally (vi) an external force $F_{a}=T_{ma}\ell ^{m}$.

\item \textit{The contraction $P^ak_a$:} The contraction with $k^{a}$, as we have mentioned,  has very interesting consequences and has not been explored adequately in the literature (except in some special cases which we will mention in the sequel). Since this contraction picks out the component  flowing along the null geodesics on the null surface (i.e the component of $P^a$ along $\ell^a$) it encodes an \textit{intrinsic property} of the null surface. It  is, therefore, worthwhile to examine this in  detail for a general case which will be the main thrust of this paper. We will show that this leads to the thermodynamic identity we are after.
\end{itemize}

The above separation of the components of $P^a$ along $(q^a_b,\ell^a,k^a)$ provides a clear picture of different aspects of gravitational dynamics on a null surface and allows us to identify which of the previous results arise from which component of $P^a$. 
\section{Thermodynamic identities from gravitational dynamics near a null surface}\label{sec:GRtherm}

We are interested in the structure of $P^ak_a$ and its interpretation as a thermodynamic identity. To study this, we begin (following \cite{Kothawala:2010bf,Hayward:1997jp}) by introducing the notion of a transverse metric $g_{ab}^{\perp}$ and the work function $P$. Let $u_{a}$ be a normalized timelike vector while $r_{a}$ be another normalized but spacelike vector related to our null vectors ($\ell_a,k_a$) by $u_{a}=(1/2A)\ell _{a}+Ak_{a}$ and $r_{a}=(1/2A)\ell _{a}-Ak_{a}$, where $A$ is an arbitrary function. Then the transverse metric defined as $g_{ab}^{\perp}=u_{a}u_{b}-r_{a}r_{b}=\ell_{a}k_{b}+\ell_{b}k_{a}$. The work function of the matter is defined \cite{Kothawala:2010bf,Hayward:1997jp} as $P=(1/2)T_{ab}g^{ab}_{\perp}=T_{ab}\ell ^{a}k^{b}$. (In the case of spherically symmetric spacetime, $P$ will be the transverse pressure; we will not bother describe the physical meaning of $P$ here since it has been done in previous literature.) When Einstein's equations hold, the work function will be 
proportional to $G_{ab}\ell ^{a}k^{b}=(1/2)G_{ab}g^{ab}_{\perp}$.

We will now study the form of equations which arise when we project the field equations along $\ell^bk^{a}$ which will lead to a  thermodynamic identity. While this can be done directly (see \ref{Paper04:AppA} for such derivation), it is nicer to obtain it from the expression for a Noether current which we will now briefly introduce. 

One can associate a natural conserved current $J^{a}=\nabla _{b}J^{ab}$ with \textit{any} vector field $v^a$ by choosing  antisymmetric second rank tensor field $J^{ab}$ corresponding to this vector field as
\begin{align}
16\pi J^{ab}=\nabla ^{a}v^{b}-\nabla ^{b}v^{a}
\end{align}
The resulting conserved current $J^a$ is indeed the standard Noether current but this approach \emph{delinks} the Noether current from diffeomorphism invariance of the action etc. and attributes the conservation law to a simple identity in differential geometry. This conserved current $J^a$, for the vector field $v^{a}$, has the following expression in general relativity \cite{Parattu:2013gwb}:
\begin{align}\label{Paper04:Sec:02:Eq.03}
16\pi J^{a}(v)=\nabla _{b}[\nabla ^{a}v^{b}-\nabla ^{b}v^{a}]=
2R^{a}_{b}v^{b}+g^{ij}\pounds _{v}N^{a}_{ij}
\end{align}
where 
\begin{equation}
N^{a}_{ij}=-\Gamma ^{a}_{ij}+(1/2)\left(\delta ^{a}_{i}\Gamma ^{k}_{kj}+\delta ^{a}_{j}\Gamma ^{k}_{ki}\right)                                                                                                                \end{equation} 
is a linear combination of Christoffel symbols. (Its physical significance is discussed in \cite{Parattu:2013gwb} and will not be repeated here).

For our purpose, we will concentrate on $16\pi k_{a}J^{a}(\xi)$, which contains the combination $R_{ab}\xi^a k^b$ (which will become $R_{ab}\ell^a k^b$ on the null surface). This is given in our coordinate system by $-16\pi J^{u}$, the component of the Noether current along the null geodesics on the surface. It can be worked out in the  most general case (presented in \ref{Paper04:AppC}), but algebraic complexity of the resulting expressions hide the physical interpretation. To bring out the physics involved, we will consider a slightly constrained situation in the main text, leaving the discussion of the most general case to \ref{Paper04:AppC}. 
The simpler case is obtained by setting (a) $\beta _{A}\vert _{r=0}=0$ just on the null surface but is arbitrary otherwise and (ii) imposing hypersurface orthogonality on the 4-velocity constructed out of the vector $\xi ^{a}$ (see \ref{app:static}). Then, from \ref{zeroth_law} in \ref{app:static}, we get the result $\partial _{A}\alpha \vert _{r=0} =0$ on the null surface. Thus, the two conditions (viz., $\beta _{A} \vert _{r=0}=0$ and hypersurface orthogonality for the 4-velocity constructed out of $\xi _{a}$) lead to the result that $\alpha$ is independent of the transverse coordinates on the null surface, which can be thought of as an extension of the zeroth law of black hole thermodynamics to a null surface in a \textit{time dependent} situation. 

In this case, the Noether current contracted with $k_{a}$ has the following expression (see \ref{Paper04:AppC} for details):
\begin{align}\label{Paper04:Sec:02:Eq.08}
16\pi k_{a}J^{a}(\xi)&=4\partial _{r}\alpha +2\alpha \partial _{r}\ln \sqrt{q}
\end{align}
However, from \ref{Paper04:Sec:02:Eq.03}, we can also rewrite the above contraction of Noether current as follows:
\begin{align}\label{LieVar}
16 \pi k_{a}J^{a}(\xi)&=2R_{ab}\xi ^{a}k^{b}+k_{a}g^{ij}\pounds _{\xi}N^{a}_{ij}
\nonumber
\\
&=2G_{ab}\xi ^{a}k^{b}-R+k_{a}g^{ij}\pounds _{\xi}N^{a}_{ij}
\end{align}
We next write the Ricci scalar $R$ in terms of the two-dimensional Ricci scalar $R^{(2)}$ for the two-surface as
\begin{align}\label{ricciGNC}
\frac{1}{2}\left(R-R^{(2)}\right)&=-2\partial _{r}\alpha -2\alpha \partial _{r}\ln \sqrt{q}
-\frac{1}{\sqrt{q}}\partial _{A}\left(\sqrt{q}\beta ^{A}\right)-\frac{3}{4}\beta ^{2}
\nonumber
\\
&+\frac{1}{4}\partial _{u}q_{AB}\partial _{r}q^{AB}+\partial _{u}\ln \sqrt{q}\partial _{r}\ln \sqrt{q}
-\frac{2}{\sqrt{q}}\partial _{u}\partial _{r}\sqrt{q}~.
\end{align}
Further, the Lie variation term in \ref{LieVar} has the following expression (see \ref{Paper04:AppC}):
\begin{equation}\label{LieVar01}
k_{a}g^{ij}\pounds _{\xi}N^{a}_{ij}=-2\pounds _{\xi}N^{u}_{ur}-q^{AB}\pounds _{\xi}N^{u}_{AB}
=-2\partial _{u}\partial _{r}\ln \sqrt{q}+\frac{1}{2}\partial _{u}q_{AB}\partial _{r}q^{AB},
\end{equation}
Thus, using \ref{ricciGNC} and \ref{LieVar01} in \ref{LieVar} and using Einstein's equations along with \ref{Paper04:Sec:02:Eq.08}, we obtain the expression for $T_{ab}\xi ^{a}k^{b}$ as
\begin{align}\label{Paper04:Sec:02:Eq.09}
-T_{ab}\xi ^{a}k^{b}=\frac{1}{8\pi}\Big(-\frac{1}{2}R^{(2)}+\alpha \partial _{\lambda}\ln \sqrt{q}
+\frac{1}{\sqrt{q}}\partial _{u}\partial _{\lambda}\sqrt{q}\Big)~.
\end{align}
In the null limit, $-T_{ab}\xi ^{a}k^{b}=T^{ab}k_{a}\xi _{b}=T^{ur}$ is the work function defined previously \cite{Hayward:1997jp}, which enables us to write the above equation on being multiplied by $\sqrt{q}$ as
\begin{align}\label{Paper04:NewAdd}
T^{ur}=\frac{1}{\sqrt{q}}\left(\frac{\alpha}{2\pi}\right)\dfrac{d}{d\lambda}\left(\frac{\sqrt{q}}{4} \right)-\frac{1}{\sqrt{q}}\left(\sqrt{q}\frac{1}{2}\frac{R^{(2)}}{8\pi}-\frac{1}{8\pi}\partial _{\lambda}\partial _{u}\sqrt{q}\right)
\end{align}
Multiplying the above equation by $\delta \lambda = \delta r$ and then integrating over a $u=\textrm{constant}$ slice of the null surface with area element $d^{2}x\sqrt{q}$, we arrive at
\begin{align}\label{Paper04:Sec:02:Eq.10}
\int d^{2}x \sqrt{q} \delta \lambda T^{ur}&=\left(\frac{\alpha}{2\pi}\right)\delta _{\lambda}\left(\int d^{2}x \frac{\sqrt{q}}{4} \right)-\delta_\lambda \left\lbrace \int d^{2}x \sqrt{q}\frac{1}{2}\frac{R^{(2)}}{8\pi}-\frac{1}{8\pi}\int d^{2}x \partial _{\lambda}\partial _{u}\sqrt{q}\right\rbrace,
\end{align} 
where we have made use of $\partial_A \alpha \vert_{r=0}$ to take $\alpha$ outside the integral and used the notation $\delta _{\lambda}=(\delta {\lambda})\partial _{\lambda}$. ($\delta_{\lambda}$ can be thought of as the change due to a virtual displacement along the vector $\delta x^a=-k^a\delta {\lambda}$. We shall explain the meaning of virtual displacement shortly.)

We take $(\alpha /2\pi)$ as the temperature of the null surface related to the surface gravity $\alpha$ and $dS=d^{2}x \sqrt{q}/4$ as the entropy associated with a proper transverse area element $d^{2}x \sqrt{q}$. Further, in the relation 
\begin{align}\label{Paper04:Sec:02:Eq.11}
\int P\sqrt{q}d^{2}x\delta \lambda=\int d^{2}x \sqrt{q} \delta \lambda~ T^{ur}=\int d^{2}x \sqrt{q} \delta \lambda~ T^{ru},
\end{align}
where $P$ is the work function,  the integral on the left hand side represents the amount of work done in a virtual displacement of the null surface by an amount $\delta \lambda$ along $k_{a}$. Then \ref{Paper04:Sec:02:Eq.10} can be recast in the following form:
\begin{align}\label{Paper04:Sec:02:Eq.12}
\bar{F}\delta \lambda =T\delta _{\lambda}S-\delta _{\lambda}E~,
\end{align}
where we have defined the energy swept out by the null surface, $\delta _{\lambda}E$, as
\begin{align}\label{Paper04:Sec:02:Eq.13}
\delta _{\lambda} E= \delta \lambda \left(\frac{\chi}{4}-\frac{\partial_{\lambda}\partial _{u}\mathcal{A}_{\perp}}{8\pi}\right)
\end{align}
with $\chi$ representing the Euler characteristic of the $2$-dimensional slice of the null surface transverse to $\ell^a$ and $k^a$,
\begin{align}
\chi = \frac{1}{4 \pi} \int \sqrt{q}d^2x R^{(2)},
\end{align}
and $\mathcal{A}_{\perp}$ representing the area of this slice. (If the 2-dimensional surface is not compact, we cannot introduce the Euler number but $\chi$ is still defined by this integral in our result.) Performing an indefinite integral along $\lambda$, this can also be written as:
\begin{equation}\label{Paper04:AppA:Eq.14aa}
E=\frac{1}{2}\int d\lambda \left(\frac{\chi}{2}\right)-\frac{1}{8\pi}\int d^{2}x\partial _{u}\sqrt{q}
\end{equation}
As an aside, we note that when $\beta_A\neq0$ we pick up a `kinetic energy' term $(1/2)\beta_A\beta^A$ in the expression for the energy and the result is given by (see \ref{Paper04:AppA:Eq.14} in \ref{Paper04:AppA})
\begin{equation}\label{Paper04:AppA:Eq.14a}
E=\frac{1}{2}\int d\lambda \left(\frac{\chi}{2}\right)-\frac{1}{8\pi}\int d^{2}x\partial _{u}\sqrt{q}-\frac{1}{16\pi}\int d \lambda \int d^{2}x \sqrt{q}\left\lbrace \frac{1}{2}\beta_A\beta^A \right\rbrace~.
\end{equation}
The notion of the virtual displacement introduced here is a straightforward generalization of the idea discussed in previous works (see e.g. \cite{Padmanabhan:2002sha,Kothawala:2009kc}). (In these earlier works, one considered spherically symmetric and static spacetimes and concentrated on the horizon as the null surface; here we have made no restrictive assumptions and deal with an arbitrary null surface.) To see the correspondence explicitly, consider the simpler situation of a static and spherically symmetric spacetime, such that $-g_{tt}=g^{rr}=f(r)$, with a (non-extremal) horizon at  $r=a$. Then $f(a)=0$ with $f'(a)$ related to the horizon temperature by $T=f'(a)/4\pi$ with `prime' denoting derivative with respect to $r$. Repeating our exercise, treating the horizon as our chosen null surface, will lead to the relation: $f'(a)a-1=8\pi P a^{2}$, where $P=T^{r}_{r}$ is the radial pressure. (The analogue of this relation in the general case of an arbitrary null surface is given by \ref{Paper04:NewAdd}). If 
we multiply this relation by $\delta a$, we can rewrite the equation in the form $T \delta S-\delta E = P \delta V$, purely algebraically. We can interpret this relation in this case by considering two solutions of the field equations differing infinitesimally such that horizons are located at $a$ and $a+\delta a$ with all other infinitesimal differences treated as the differences between these two solutions. (This is analogous to the relation for general null surface given in \ref{Paper04:Sec:02:Eq.12}). Hence, the virtual displacement is essentially a shift between the location of the fiducial null surface in two solutions of field equations. The shift  moves the null surface by an amount $\delta \lambda$, where $\lambda$ is the affine parameter along the null geodesics of $k^{a}$. More detailed discussion of this idea can be found in \cite{Kothawala:2009kc,Kothawala:2007em,Padmanabhan:2009vy,gravitation}.

Coming back to the case with $\beta_A=0$, we note that $\bar{F}$ represents the integral of the work function over the null surface i.e. $\bar{F}=\int P\sqrt{q}d^{2}x$. Then \ref{Paper04:Sec:02:Eq.12} is better interpreted when rewritten in the form
\begin{align}\label{Term_identity}
\delta_{\lambda}{E} =T\delta _{\lambda}S-\bar{F}\delta \lambda~.
\end{align}
This expression is quite suggestive. The virtual displacement can be interpreted as a physical process that displaced the null surface from $r=0$ to $r=r+\delta \lambda$. Then, the energy engulfed by the null surface in this displacement is the sum of a heat energy (viz. the temperature multiplied by the change in entropy) and the work done during this virtual displacement of the null surface. The above equation can also be interpreted as the total energy in the region being a sum of an energy corresponding to matter, represented by the work done term, and energy corresponding to pure gravity, represented by the heat term. This is the most general form of the thermodynamic identity which arises from the projection of $R_{ab}\ell^a$ along $k^a$. We shall now discuss the applications of this result to special cases.
\section{Special Cases}\label{sec:special}

In the previous section, we have shown the equivalence of gravitational field equations with a thermodynamic identity for an arbitrary null surface. The result in \ref{sec:GRtherm} has been obtained  in the earlier literature  for some special cases.  In this section, we will connect up with the earlier work by specializing this result to (i) stationary  spacetimes without any other symmetry
(ii) spherically symmetric spacetime which is not necessarily static and (iii) static spherically symmetric spacetime.
\subsection{Stationary spacetime}\label{on_null_stationary}

Since we have identified $\xi^a$ as our time flow vector, stationarity involves setting partial derivatives of metric components with respect to $u$ to zero (see \ref{app:static}). In this case the thermodynamic identity in \ref{Paper04:Sec:02:Eq.12} retains its form with a simpler expression for the  energy term. The expression for energy in \ref{Paper04:Sec:02:Eq.13} becomes
\begin{align}\label{New_Ref}
\delta _{\lambda} E&= \frac{\chi}{4} \delta \lambda= \frac{\chi}{4} \delta r 
\end{align}
This immediately implies
\begin{align}
\frac{\partial E}{\partial r} &=\frac{\chi}{4}~.
\end{align}
This matches with the result in \cite{Kothawala:2009kc}. Notice that, even in the more general case (when $E$ is given by \ref{Paper04:AppA:Eq.14a}), we can obtain the same result if we assume (a)$\beta_A=0$ on the null surface and (b) the stationarity condition, viz. the metric is independent of $u$. So the above result does not require the spacetime to be static (which involves the additional condition of hypersurface orthogonality) but only requires stationarity.

The additional restrictions required for achieving staticity are the conditions of hypersurface-orthogonality given in \ref{HSO1} and \ref{HSO2} in \ref{app:static}, reproduced below:
\begin{align}
r \left(\beta_2 \partial_r \beta_1 -\beta_1 \partial_r \beta_2\right) -  \left(\partial_1 \beta_2-\partial_2 \beta_1\right)&=0~,\label{HSO11} \\
r \alpha \partial_r \beta_A -  \partial_A \alpha -r \beta_A \partial_r \alpha &=0\label{HSO21}~.  
\end{align}  
As we have noted in \ref{app:static}, these conditions would imply $\partial_A \alpha \vert_{r=0}=0$. 
There is no modification to our result above since we had already assumed hypersurface-orthogonality.
\subsection{Spherically symmetric spacetime}

To restrict the GNC metric in \ref{Paper04:Sec:02:Eq.01} to a spherically symmetric form, the most convenient way would be to enforce the geometry of $2$-spheres on the $u=$constant, $r$=constant $2$-surfaces. However, the $u=$constant surfaces should not be considered as constant-time surfaces as these surfaces are actually null. Thus, identifying $x^A$ with the angular coordinates, we should demand $\partial_A \alpha=0$, $\beta^A=0$ and $q_{AB}= f(u,r) d \Omega^2 = f(u,r) \left(d\theta^2 + \sin^2 \theta d \phi^2\right)$ to arrive at a spherically symmetric form of GNC. The form of the line element now becomes
\begin{equation}
ds^2= -2r \alpha(r,u)du^2 + 2 du dr +f(u,r) d \Omega^2,
\end{equation}
which is of spherically symmetric form \cite{LL2}. The $2$-surface we are mainly interested is $u=$constant cross-section of the null surface at $r=0$. Defining the ``radial coordinate'' \cite{Wald}, $R(u,r)\equiv \sqrt{f(u,r)}$, and expanding in Taylor series in $r$ around $r=0$, we obtain $R(u,r)=R_{H}(u)+rg(r,u)$, where the last term is not just the linear order term in the Taylor expansion but represents all the higher order terms taken together. The null surface at a constant $u$ has a ``radius'' $R_{H}(u)$, with the $u$-dependence allowing for the area of the $2$-surface to be changing with $u$. (So we have assumed spherical symmetry but have allowed for time dependence).

Again, our result holds with a simpler expression for the energy. To see this, let  us  look at \ref{Paper04:Sec:02:Eq.13}. For $2$-spheres, the Euler characteristic equals $2$. Substituting in \ref{Paper04:Sec:02:Eq.13}, we obtain
\begin{equation}
 \delta _{\lambda} E= \delta \lambda \left(\frac{1}{2}-\frac{\partial_{\lambda}\partial _{u}\mathcal{A}_{\perp}}{8\pi}\right),
\end{equation}
where $\mathcal{A}_{\perp}$ is the area of the compact $2$-surface. Interpreting $E$ as the energy associated with a $u=$constant, $r$=constant $2$-surface, we have $E=E(u,r)$ for the spherically symmetric spacetime. The above equation can then be written as a partial differential equation as follows:
\begin{equation}
 \frac{\partial E}{\partial \lambda} = \frac{1}{2}-\frac{\partial_{\lambda}\partial _{u}\mathcal{A}_{\perp}}{8\pi},
\end{equation}
where $\partial/\partial \lambda$ is the same as $\partial/\partial r$, since $r=\lambda-\lambda_H$, and is taken keeping $u$ constant. The solution is
\begin{equation}\label{E_gen}
 E(\lambda,u)=\frac{\lambda}{2}-\frac{\partial _{u}\mathcal{A}_{\perp}}{8\pi}+F(u),
\end{equation}
where $F$ is an arbitrary function. In order to fix this function, let us consider the ingoing null geodesics, $-\partial/\partial r$ on the $u=$constant surface. These are $\theta=$constant, $\phi=$constant lines and are hence radial. Let us assume that moving along this geodesic will lead us to intersect ever smaller two-spheres, i.e we will be moving towards the inner part of the $2$-spheres. (If this is not the case, we can move in the other direction.) Due to spherical symmetry, all these geodesics will intersect at the common center of the $2$-spheres (assuming that the geometry is such that this center exists). Since all the geodesics are given the affine parameter value $\lambda_H$ at the $r=0$ $2$-sphere, they will all have the same value of affine parameter $\lambda$ (or, equivalently, $r$), at the center. But this affine parameter value may be different on a different $u$-constant surface. Let us label this value $\lambda_{0}(u)$. At the center, \ref{E_gen} becomes
\begin{equation}\label{E_cen}
 E(\lambda_{0}(u),u)=\frac{\lambda_{0}(u)}{2}-\frac{\partial _{u}\mathcal{A}_{\perp}}{8\pi}\vbar_{\lambda=\lambda_{0}(u)}+F(u),
\end{equation}
Now, $\mathcal{A}_{\perp} = 4 \pi f = 4 \pi R^2$ so that $\partial _{u}\mathcal{A}_{\perp}= 8 \pi R \partial_u R$, which is zero at $\lambda=\lambda_{0}(u)$ since $R=0$. Thus, we obtain
\begin{equation}\label{E_cen1}
 E(\lambda_{0}(u),u)=\frac{\lambda_{0}(u)}{2}+F(u).
\end{equation}
Since it seems natural to associate zero energy to a $2$-sphere that is essentially a point, we shall choose $F(u)=-\lambda_{0}(u)/2$. Thus, our definition for energy becomes
\begin{equation}\label{E_spec}
 E(\lambda,u)=\frac{\lambda-\lambda_{0}(u)}{2}-\frac{\partial _{u}\mathcal{A}_{\perp}}{8\pi}~.
\end{equation}
For the $r=0$ surface, we have the energy as
\begin{align}
 E(\lambda_H,u)&=\frac{\lambda_H-\lambda_{0}(u)}{2}-\frac{\partial _{u}\mathcal{A}_{\perp}\vbar_{r=0}}{8\pi} \nn\\
 &=\frac{\lambda_H-\lambda_{0}(u)}{2}-R_H \partial_u R_H~. \label{E_hor}
\end{align}
A special case of interest is the case in which the $2$-sphere line element is $(r+R_{H})^2 d\Omega^2$, for a constant $R_{H}$, which puts $R=r+R_H$. Taking $R$ as a coordinate instead of $r$, we obtain the metric element 
\begin{equation}\label{GNC-spec}
 ds^{2}=-2(R-R_H)\alpha du^{2}+2dudR+R^2 d\Omega^2,
\end{equation}
which is of the form of the metric element used in \cite{Papnoi:2014swa}. (The metric that the authors of \cite{Papnoi:2014swa} start with is different in form, but they reduce it to this form through a coordinate transformation.) In this case, $\partial_u R_H=0$, $\lambda_H-\lambda_{0}(u)=r_H - r_{0}(u)= R_{H}-R_0=R_{H}$, where we have used $r=\lambda-\lambda_H$, denoted $r$ at our fiducial null surface as $r_H$, even though it is zero in our framework, and the $r$ and $R$ at the center of the $2$-spheres as $r_{0}$ and $R_{0}$ respectively. Substituting, we get the energy of a $u=$constant, $r=0$ surface as
\begin{align}
 E(\lambda_H,u)&=\frac{R_H}{2} \label{E_hor_spec}
\end{align}
matching with the results of \cite{Papnoi:2014swa}. However, we should stress that they have derived the first law for an apparent horizon which is \textit{not} a null surface while in this work we have derived a similar  result for a null surface. 

With hindsight, it may have been better to set the origin of the affine parameter at the center of the $2$-spheres, i.e $\lambda_{0}(u)$, to obtain a form for the energy as
\begin{equation}
 E(\lambda_H,u)=\frac{\lambda_H(u)}{2}-R_H \partial_u R_H~, \label{E_alt}
\end{equation}
where $\lambda_H$, the value of the affine parameter at $r=0$, is no longer a constant but depends on $u$.
\subsubsection{Stationary spherically symmetric case}

Since $\beta^A=0$ and $\partial_A \alpha=0$, enforcing the stationary conditions $\partial_u \alpha=0$ and $\partial_u f=0$ leads to staticity (see \ref{HSO11} and \ref{HSO21}). This is not surprising as every stationary spherically symmetric spacetime is automatically static\cite{Wald}. Thus, the previous results for static spacetime hold with the spherically symmetric transverse metric. In this case \ref{E_hor} leads to
\begin{align}
E=\frac{\lambda_H-\lambda_{0}(u)}{2},
\end{align}
while using \ref{E_alt} leads to
\begin{align}
 E(\lambda_H,u)&=\frac{\lambda_H}{2}. \label{E_alt1}
\end{align}
which matches with previous results \cite{Kothawala:2009kc}. Note that the result in \cite{Kothawala:2009kc} was obtained by setting an arbitrary integration constant to zero to ensure $E \rightarrow 0 $ when $\lambda_H=0$. In the spherically symmetric case, this has a very physical interpretation as the radius of the $2$-sphere shrinks to zero.
\section{Discussion}

We started this work trying to address the question of whether the gravitational field equations near \textit{any  null surface in an arbitrary spacetime} reduces to a thermodynamic identity, generalizing results previously available in the literature for special cases.  We have shown in \ref{sec:GRtherm} that this is indeed possible, by introducing (a) the temperature  through surface gravity, (b) entropy density from the area and (c) the work function from the transverse metric $g_{ab}^{\perp}$.
We then obtain, by projecting the Einstein's equations along the $k^a$ direction, a relation of the form $T\delta_\lambda S=\delta_\lambda E+P\delta_\lambda V$ where the variations represent virtual displacements of the null surface along null geodesics off the surface. 

Given an arbitrary null surface with associated normal $(\ell^a)$, co null vector $(k^a)$ and the transverse metric ($q^a_b$), one can study the projections of the vector $P^a\propto R^a_b\ell^b$, along each of these. 
We  pointed out that the projection of $P^a$ along $q_{a}^{b}$ leads to the Navier-Stokes equation on the null surface while the projection along $\ell^a$ is related to the Raychaudhuri equation. This clearly shows that all the information contained in the field equations posses thermodynamic interpretation. 

As an aside, note that our result, arising from projection $P^ak_a$ along $k^a$, is distinct from any result (like e.g the connection with Clausius relation \cite{Jacobson:1995ab}) obtained from projection $P^a\ell_a$ along $\ell^a$ (and the resulting the Raychaudhuri equation) and \textit{these two class of results should not be confused with each other}. As noted earlier, the results based on projection along $\ell^a$ (viz. the Raychaudhuri equation) could not be generalized in a simple manner to \LL\ models and they need additional restrictive assumptions (like the vanishing of shear and expansion) even for thermodynamic interpretation. In fact, as the earlier work in \cite{Kothawala:2010bf} clearly points out, the thermodynamic structure of the curvature tensor is not properly captured in the components which occur in the projection along $\ell^a$ and the  Raychaudhuri equation. We believe the other two projections (on $q_{ab}$ and on $k_a$) leads to richer thermodynamic content. This is because they 
pick out the components of $P^a$ along $\ell^a$ (due to $\ell^2=0,\ell^ak_a=-1$!) and along $e^a_A$ both of which are intrinsic to the null surface.  They  are also most likely to remain valid even in a more general class of theories. (We already know that if the spacetime is static, then the resulting thermodynamic identity holds even for \LL\ models.). This clarification of the different projections of $R^a_b\ell^b$ and their thermodynamic relevance is an important offshoot of our work.

We derived our result starting from the Noether current, which shows again the intimate connection between Noether charge and thermodynamics seen in earlier works. Through this exercise, we have introduced a definition of energy which reduces to 
energy definitions introduced previously in the static case \cite{Kothawala:2009kc}. In the most general context, this involves time derivatives of the area of the null surface and additional terms involving off-diagonal metric elements $\beta _{A}$. If we assume $\beta _{A}=0$ on the null surface, enforcing hypersurface-orthogonality on our chosen time-flow vector naturally leads to $\partial _{A}T\vert _{r=0}=0$, which is an extension of the zeroth law of thermodynamics to the case of an arbitrary null surface. In this situation, the energy consists of two parts: the standard two dimensional curvature scalar, related to the Euler characteristic of the null surface, and  a term involving time rate of change of the null surface area. Since the two-metric is independent of time in the static case, energy becomes solely dependent on the Euler characteristic of the null surface. We then discuss the case of stationary, static, spherically symmetric and stationary spherically symmetric spacetimes and make 
connection with results previously available in the literature.

To summarize, we have shown that for any arbitrary spacetime, without assuming any symmetry, gravitational field equations in general relativity near an arbitrary null surface reduces to a thermodynamic identity. Also, for a restricted class of spacetimes with hypersurface orthogonality enforced, zeroth law holds (even in \emph{time dependent} cases). It is interesting to ask whether identical results hold for the most general class of gravitational Lagrangians with second order equations of motion, i.e. in \LL gravity, as well since the previous result for static spacetime was indeed applicable to \LL\ models. This work is under progress  and the results will be presented elsewhere. 

\section*{Acknowledgement}

Research of T.P is partially supported by J.C. Bose research grant of DST, Govt. of India. Research of S.C and K.P. are supported by SPM Fellowship from CSIR, Govt. of India. The authors also thank Suprit Singh and Kinjalk Lochan for helpful discussions.

\appendix
\labelformat{section}{Appendix #1} 

\section*{Appendices}
\section{Derivation from gravitational field equations}\label{Paper04:AppA} 
In a static and spherically symmetric spacetimes, we have the relation $G^{t}_{t}=G^{r}_{r}$ between the Einstein tensor components in the near horizon limit. This relation continues to hold for arbitrary static spacetimes as well \cite{Medved:2004ih}. Taking a cue from this, let us evaluate the corresponding Einstein tensor components for the GNC coordinates. All expressions are evaluated on the null surface, i.e at $r=0$. We have
\begin{align}\label{Paper04:AppA:Eq.04}
G^{u}_{u}&=g^{ua}G_{au}=G_{ur};\qquad G^{r}_{r}=g^{ra}G_{ar}=G_{ur}
\nonumber
\\
G^{u}_{u}&=G^{r}_{r}=G_{ur}=R_{ur}-\frac{1}{2}R
=R_{ur}-\frac{1}{2}\left(2R_{ur}+\mu ^{AB}R_{AB}\right)
=-\frac{1}{2}q^{AB}R_{AB}
\nonumber
\\
&=\frac{1}{2}\alpha q^{AB}\partial _{r}q_{AB}-\frac{1}{2}R^{(2)}+\frac{1}{2}q^{AB}\partial _{r}\partial _{u}q_{AB}+\partial _{r}\ln \sqrt{q}\partial _{u}\ln \sqrt{q}+\frac{1}{2}\partial _{u}q_{AB}\partial _{r}q ^{AB}
\nonumber
\\
&+\frac{1}{4}\left(\beta ^{A}\beta _{A}+\beta ^{A}q^{CD}\partial _{A}q_{CD}
-2\beta ^{A}q^{CD}\partial _{C}q_{AD}+2q^{AB}\partial _{A}\beta _{B} \right)
\nonumber
\\
&=\alpha \partial _{r}\ln \sqrt{q}-\frac{1}{2}R^{(2)}
+\frac{1}{2}q^{AB}\partial _{r}\partial _{u}q_{AB}
+\partial _{r}\ln \sqrt{q}\partial _{u}\ln \sqrt{q}
+\frac{1}{2}\partial _{u}q_{AB}\partial _{r}q^{AB}
\nonumber
\\
&+\frac{1}{4}\left(\beta ^{A}\beta _{A}+\beta ^{A}q^{CD}\partial _{A}q_{CD}
-2\beta ^{A}q^{CD}\partial _{C}q_{AD}+2q^{AB}\partial _{A}\beta _{B} \right)
\end{align} 
Also, we have the following identity:
\begin{align}\label{Paper04:AppA:Eq.05}
-\frac{1}{2}q^{AB}R_{AB}&=-\frac{1}{2}\delta ^{A}_{B}R^{B}_{A}
=-\frac{1}{2}\delta ^{A}_{B}\left(R^{Bu}_{Au}+R^{BC}_{AD}\delta ^{D}_{C}+R^{Br}_{Ar}\right)
\nonumber
\\
&=-\delta ^{A}_{B}R^{Bu}_{Au}-\frac{1}{4}\delta ^{AB}_{CD}R^{CD}_{AB},
\end{align}
where $\delta ^{AB}_{CD}=\delta ^{A}_{C}\delta ^{B}_{D}-\delta ^{A}_{D}\delta ^{B}_{C}$ and we have used the relation
\begin{align}\label{Paper04:AppA:Eq.06}
\delta ^{A}_{B}R^{Br}_{Ar}&=\delta ^{A}_{B}R^{B}_{uAr}=q^{AB}R_{BuAr}
\nonumber
\\
&=q^{AB}R_{BuAr}=q^{AB}R_{ArBu}=\delta ^{B}_{A}R^{Au}_{Bu}
\end{align}
The terms in \ref{Paper04:AppA:Eq.04} involving $q_{AB}$ and its derivative can be manipulated leading to
\begin{align}\label{Paper04:AppA:Eq.07}
\frac{1}{2}q^{AB}\partial _{r}\partial _{u}q_{AB}&+\partial _{r}\ln \sqrt{q}\partial _{u}\ln \sqrt{q}+\frac{1}{2}\partial _{u}q_{AB}\partial _{r}q^{AB}
\nonumber
\\
&=\frac{1}{2}\partial _{r}\left(q^{AB}\partial _{u}q_{AB}\right)
-\frac{1}{2}\partial _{r}q^{AB}\partial _{u}q_{AB}+\partial _{r}\ln \sqrt{q}\partial _{u}\ln \sqrt{q}+\frac{1}{2}\partial _{u}q_{AB}\partial _{r}q^{AB}
\nonumber
\\
&=\partial _{r}\left(\frac{\partial _{u}\sqrt{q}}{\sqrt{q}}\right)+\frac{\partial _{r}\sqrt{q}\partial _{u}\sqrt{q}}{\left(\sqrt{q}\right)^{2}}=\frac{1}{\sqrt{q}}\partial _{r}\partial _{u}\sqrt{q}
\end{align}
The terms with $\beta _{A}$ in \ref{Paper04:AppA:Eq.04} can be simplified leading to
\begin{align}\label{Paper04:AppA:Eq.08}
\frac{1}{4}\Big(\beta ^{A}\beta _{A}&+\beta ^{A}q^{CD}\partial _{A}q_{CD}
-2\beta ^{A}q^{CD}\partial _{C}q_{AD}+2q^{AB}\partial _{A}\beta _{B} \Big)
=\frac{1}{4}\beta ^{2}+\frac{1}{2}\beta ^{A}\partial _{A}\ln \sqrt{q}
\nonumber
\\
&-\frac{1}{2}\partial _{C}\beta ^{C}+\frac{1}{2}q_{AD}\partial _{C}\left(\beta ^{A}q^{CD}\right)
+\frac{1}{2}q^{AB}\partial _{A}\beta _{B}
\nonumber
\\
&=\frac{1}{4}\beta ^{2}+\frac{1}{2\sqrt{q}}\partial _{A}\left(\sqrt{q}\beta ^{A}\right)
\end{align}
Then, substituting all the expressions in \ref{Paper04:AppA:Eq.04}, we arrive at the following result:
\begin{align}\label{Paper04:AppA:Eq.09}
G^{r}_{r}&=-\delta ^{A}_{B}R^{uB}_{uA}-\frac{1}{4}\delta ^{AB}_{CD}R^{CD}_{AB}=-\frac{1}{2}q^{AB}R_{AB}
\nonumber
\\
&=\alpha \partial _{r}\ln \sqrt{q}-\frac{1}{2}R^{(2)}
+\frac{1}{\sqrt{q}}\partial _{r}\partial _{u}\sqrt{q}+\frac{1}{4}\beta ^{2}
+\frac{1}{2\sqrt{q}}\partial _{A}\left(\sqrt{q}\beta ^{A}\right)
\end{align}
For displacement of the null surface by an amount $\delta \lambda$ along the ingoing null geodesic, we multiply the above equation by $\delta \lambda \sqrt{q}$ and use the gravitational field equation $G^{r}_{r}=8\pi T^{r}_{r}$, leading to
\begin{equation}\label{Paper04:AppA:Eq.10}
8\pi T^{r}_{r}\delta \lambda \sqrt{q} =\alpha \delta _{\lambda}\sqrt{q}-\frac{1}{2}R^{(2)}\delta \lambda \sqrt{q}+\delta \lambda \partial _{\lambda}\partial _{u}\sqrt{q}+\left[\frac{1}{4}\beta ^{2}
+\frac{1}{2\sqrt{q}}\partial _{A}\left(\sqrt{q}\beta ^{A}\right)\right]\delta \lambda \sqrt{q}
\end{equation}
where we have used the relation, $\delta _{\lambda}f=(\partial f/\partial \lambda)\delta \lambda$, for any scalar function $f$. Then, dividing the above equation by $8\pi$ and integrating over a two dimensional surface with $d^{2}x$ we arrive at
\begin{align}\label{Paper04:AppA:Eq.11}
\int d^{2}x \sqrt{q} \delta \lambda T^{r}_{r}&=\int d^{2}x \left(\frac{\alpha}{2\pi}\right)
\delta _{\lambda}\left(\frac{\sqrt{q}}{4} \right)-\delta \lambda \Big\lbrace \int d^{2}x \sqrt{q}\frac{1}{2}\frac{R^{(2)}}{8\pi}-\frac{1}{8\pi}\int d^{2}x \partial _{\lambda}\partial _{u}\sqrt{q}
\nonumber
\\
&-\int d^{2}x\sqrt{q}\frac{1}{8\pi}\left[\frac{1}{4}\beta ^{2}
+\frac{1}{2\sqrt{q}}\partial _{A}\left(\sqrt{q}\beta ^{A}\right)\right]\Big\rbrace
\end{align} 
The null rays tangent to the null surface have the non-affinity coefficient $\alpha$, which suggests defining $(\alpha /2\pi)$ as the temperature of the null surface. Along with this, we can interpret $T^{r}_{r}$ as the normal pressure $P_{\perp}$ \emph{on the null surface}. This identification allows us to interpret the object
\begin{equation}\label{Paper04:AppA:Eq.12}
\bar{F}=\int d^{2}x \sqrt{q} P_{\perp}
\end{equation}
as the average normal force over the \emph{null surface}. Then, $\bar{F}d\lambda$ can be interpreted as the virtual work done in displacing the null surface by $\delta \lambda$ along ingoing null geodesics. \ref{Paper04:AppA:Eq.11} can now be written as
\begin{equation}\label{Paper04:AppA:Eq.13}
\bar{F}\delta \lambda =\int ~d^{2}x T\delta _{\lambda}s-\delta _{\lambda}E,
\end{equation}
where $s$ is the entropy density of the \emph{null surface} with the following expression: $s=(\sqrt{q}/4)$, which equals the Bekenstein-Hawking entropy density. We have also identified the energy $E$ associated with the null surface as
\begin{equation}\label{Paper04:AppA:Eq.14ab}
E=\frac{1}{16\pi}\int d\lambda \int d^{2}x\sqrt{q}~R^{(2)}-\frac{1}{8\pi}\int d^{2}x\partial _{u}\sqrt{q}-\frac{1}{16\pi}\int ^{\lambda}\delta \lambda \int d^{2}x \sqrt{q}\left\lbrace \frac{1}{2}\beta ^{2}+\frac{1}{\sqrt{q}}\partial _{A}\left(\sqrt{q}\beta ^{A}\right) \right\rbrace 
\end{equation}
When the two-dimensional surface is compact, this reduces to a simpler form, given by
\begin{equation}\label{Paper04:AppA:Eq.14}
E=\frac{1}{2}\int d\lambda \left(\frac{\chi}{2}\right)-\frac{1}{8\pi}\int d^{2}x\partial _{u}\sqrt{q}-\frac{1}{16\pi}\int ^{\lambda}\delta \lambda \int d^{2}x \sqrt{q}\left\lbrace \frac{1}{2}\beta_A\beta^A \right\rbrace 
\end{equation}
where $\chi$ represents the Euler characteristic of a two-dimensional compact manifold $\mathcal{M}_{2}$ without boundary and is given by the following expression:
\begin{align}\label{Paper04:AppA:Eq.15}
\chi \left(\mathcal{M}_{2}\right)=\frac{1}{4\pi}\int _{\mathcal{M}_{2}}d^{2}x\sqrt{q}~R^{(2)}
\end{align}
Note that in this most general situation the first law has to be interpreted as follows: under infinitesimal shift of the null surface along ingoing null geodesics, change in energy and work done due to pressure adds up and yield $T\delta _{\lambda}s$ integrated over the null surface. Thus, in this general case $T\delta _{\lambda}s$ has to be interpreted locally as being due to displacement of a small element on the null surface. This difficulty arises since the temperature $\alpha/2\pi$ in the $Tds$ term is dependent on the transverse coordinates and cannot be taken outside the transverse integral. 
 The above discussion outlines the derivation of first law from the field equation perspective and matches exactly with the one obtained from Noether current formalism in \ref{Paper04:Sec:02}.

\section{GNC metric in static form}\label{app:static}

The GNC line element is 
\begin{equation}\label{GNCline}
	ds^{2}=-2r\alpha du^{2}+2drdu-2r\beta _{A}dudx^{A}+q _{AB}dx^{A}dx^{B}~.
\end{equation}
We shall attempt to reduce \ref{GNCline} to the form of the static metric in \cite{Medved:2004ih}, 
\begin{equation} \label{staticline}
	ds^{2}=-N^{2}dt^{2}+dn^{2}+\sigma _{AB}dy^{A}dy^{B}~,
\end{equation}
using appropriate restrictions and coordinate transformations. We shall place the first restriction on $\alpha$, demanding it to be positive in the region $r>0$. The utility of this restriction will be clear in due course.

\ref{GNCline} represents the line element near an arbitrary null surface in an arbitrary spacetime. To get to \ref{staticline}, we need to enforce staticity. A static spacetime should satisfy the following two requirements\cite{Wald}:
\begin{enumerate}[i)]
	\item There must exist a timelike vector $\xi^{a}$ that satisfies the Killing condition, i.e. 
	\begin{equation}\label{Kill_cond}
		\nabla_{a} \xi_{b}+\nabla_{b} \xi_{a}=0~.
	\end{equation}
	\item $\xi^{a}$ must be hypersurface-orthogonal. By Frobenius theorem, this is equivalent to demanding 
	\begin{equation}\label{Frobenius}
		\xi_{[a} \nabla_{a} \xi_{b]}=0.  
	\end{equation}
\end{enumerate}
If only the first condition holds, then the spacetime is called stationary. For stationarity, it is not necessary that the vector $\xi^{a}$ be timelike everywhere in the spacetime. If we impose that $\xi^{a}$ be timelike everywhere, then even the Schwarzschild spacetime with the Killing vector $\xi^a = \partial/\partial t$ will not be stationary. Thus, we will only demand that $\xi^{a}$ be timelike in possibly only in part of the spacetime (see \cite{Carroll}).

Since the line element \ref{GNCline} has been constructed in a region near the null surface and no claim has been made about its validity for the entire spacetime, we shall adapt the above criteria to our situation by calling a GNC metric as static if we can find a timelike vector $\xi^{a}$ that satisfies \ref{Kill_cond} and \ref{Frobenius} in the region of validity of \ref{GNCline}. We shall further restrict the domain of validity to the $r>0$ region, where $g_{uu}<0$ for $\alpha>0$, since even in a Schwarzschild spacetime the timelike Killing vector is timelike only outside the horizon. 

The next logical step would be to choose a timelike vector $\xi^{a} $ in the chosen domain and demand that it satisfies \ref{Kill_cond} and \ref{Frobenius}. While these two conditions are enough to render the spacetime static, the static line element in \ref{staticline} also has a Killing horizon at $n=0$. In other words, the norm of the Killing vector vanishes at $n=0$. We would like our null surface at $r=0$ to go to the Killing horizon in the static limit. The Killing vector for \ref{staticline} lies on the Killing horizon. Thus, we are looking for a vector $\xi^a$ that is timelike in the region $r>0$, is null at $r=0$ and lies on the null surface $r=0$. An obvious choice is the vector $\mathbf{\xi} = \partial/\partial u$. 

To strengthen the motivation for this choice, we shall now demonstrate that it corresponds to the timelike Killing vector in Schwarzschild and Rindler metrics. Both Schwarzschild and Rindler metrics have the form of the $f(r)$-metric:
\begin{equation}\label{fr_line}
	ds^2= -f(r) dt^2 + \frac{dr^2}{f(r)}+ q_{AB}dx^A dx^B~,
\end{equation}
with $f(r)=1-2M/r$ giving Schwarzschild and $f(r)=-2 \kappa r$ giving Rindler. The timelike Killing vector in the coordinate order $(t,r,x^1,x^2)$ is $\xi^{a}=\left(1,0,0,0\right)$. Defining a new coordinate $u$ by the relation
\begin{equation}
	u= t + \int \frac{dr}{f(r)},
\end{equation}
we have 
\begin{equation}
	du= dt + \frac{dr}{f(r)} \Longrightarrow dt= du - \frac{dr}{f(r)}~.
\end{equation}
Substituting in \ref{fr_line}, we obtain
\begin{equation}
	ds^2= -f(r) du^2 + 2 du dr + q_{AB}dx^A dx^B~,
\end{equation}
which is the GNC line element \ref{GNCline} with $\beta^A=0$. In coordinates $(u,r,x^1,x^2)$, $\xi^{a}=\left(1,0,0,0\right)=\partial /\partial u$, our chosen timelike Killing vector for GNC.

Having chosen a $\xi^a$, we shall now apply the conditions \ref{Kill_cond} and \ref{Frobenius}. The Killing condition gives
\begin{align}
	\pounds_{\xi}g_{ab}=0 \Rightarrow \xi^c \partial_c g_{ab} + g_{cb} \partial_a \xi^c + g_{ac} \partial_b \xi^c=0 \Rightarrow \xi^c \partial_c g_{ab}= \partial_u g_{ab}=0~.
\end{align}
Thus, the Killing condition demands that all the metric components be independent of $u$. This means
\begin{equation}\label{Killing-2}
	\partial_u \alpha =0;\quad \partial_u \beta_{A} =0; \quad \partial_u q_{AB} =0~.
\end{equation}
Next, let us look at \ref{Frobenius}. The equation $\xi_{[a} \nabla_{a} \xi_{b]}=0$ gives four equations corresponding to $(a,b,c)=(u,x^1,x^2)$, $(a,b,c)=(r,x^1,x^2)$ and two equations with $(a,b,c)=(u,r,x^A)$ for $A=1,2$. These correspond, respectively, to 
\begin{align}
	2 r^2  \alpha \left( \partial_1 \beta_2 -\partial_2 \beta_1\right)+ r^2 \left( \beta_2 \partial_u \beta_1-\beta_1 \partial_u \beta_2\right)+ 2 r^2 \left( \beta_1 \partial_2 \alpha-\beta_2 \partial_1 \alpha\right)&=0~, \label{eq1}\\
	r^2 \left(\beta_2 \partial_r \beta_1 -\beta_1 \partial_r \beta_2\right) - r \left(\partial_1 \beta_2-\partial_2 \beta_1\right)&=0~, \\
	2 r^2 \alpha \partial_r \beta_A - 2r \partial_A \alpha + r\partial_u \beta_A -2r^2 \beta_A \partial_r \alpha &=0~.  
\end{align}   
Note that these equations are not all independent. For example, enforcing the last two equations, the first one can be seen to be satisfied identically. Thus, we just need to demand
\begin{align}
	r^2 \left(\beta_2 \partial_r \beta_1 -\beta_1 \partial_r \beta_2\right) - r \left(\partial_1 \beta_2-\partial_2 \beta_1\right)&=0~, \\
	2 r^2 \alpha \partial_r \beta_A - 2r \partial_A \alpha + r\partial_u \beta_A -2r^2 \beta_A \partial_r \alpha &=0~.  
\end{align}
These equations are automatically satisfied at $r=0$. Elsewhere in the spacetime region under consideration, we can cancel a factor of a power of $r$ to get conditions on the metric components. Since we are considering smooth functions, we should expect these conditions to hold even at $r=0$. Thus, canceling off overall factors of constants and powers of $r$,  we obtain the following conditions for hypersurface-orthogonality:
\begin{align}
	r \left(\beta_2 \partial_r \beta_1 -\beta_1 \partial_r \beta_2\right) -  \left(\partial_1 \beta_2-\partial_2 \beta_1\right)&=0~, \\
	2 r \alpha \partial_r \beta_A - 2 \partial_A \alpha + \partial_u \beta_A -2r \beta_A \partial_r \alpha &=0~.  
\end{align}
Specializing to the stationary case, we enforce \ref{Killing-2} and obtain
\begin{align}
	r \left(\beta_2 \partial_r \beta_1 -\beta_1 \partial_r \beta_2\right) -  \left(\partial_1 \beta_2-\partial_2 \beta_1\right)&=0~,\label{HSO1} \\
	r \alpha \partial_r \beta_A -  \partial_A \alpha -r \beta_A \partial_r \alpha &=0\label{HSO2}~.  
\end{align}   
In particular, \ref{HSO2} implies
\begin{eqnarray} \label{zeroth_law}
	\partial_A \alpha |_{r=0} =0~.
\end{eqnarray}
Thus, imposing staticity on the GNC metric with $\mathbf{\xi} =\mathbf{\partial/\partial u}$, we get a generalization of the zeroth law of black hole thermodynamics.

Once staticity is imposed, it is advantageous to transform to a coordinate where it is manifest. Let us take the hypersurfaces to which $\xi^a$ is orthogonal to be level surfaces of a function $t$, i.e we shall take $\xi^a$ to be orthogonal to $t=\textrm{constant}$ surfaces. Then, we should have
\begin{equation} \label{eq2}
 \xi_a = F(u,r,x^A) \nabla_a t~.
\end{equation}
We shall show that there exists a $t$ which satisfies this equation if we take $F(u,r,x^A)= - 2 r \alpha$. With this choice of $F$, \ref{eq2} becomes $g_{au} = -2 r \alpha \nabla_a t $, where $\nabla_a t = -(g_{au}/2 r \alpha)$. Hence the components of the vector $\nabla _{a}t$ in GNC coordinate reads,
\begin{align}
 \left(\partial_u t, \partial_r t, \partial_A t \right) = -\frac{1}{2 r \alpha} \left(-2 r \alpha,1, -r \beta_A \right) 
\end{align}
which immediately leads to an expression for $dt$ as:
\begin{equation}
dt = du - \frac{dr}{2r\alpha}+ \frac{\beta_A dx^A}{2 \alpha}~. \label{dt_exp}
\end{equation}
For $dt$ to be a perfect differential, the following integrability conditions need to be satisfied:
\begin{align}
\partial_A \left( -\frac{1}{2 r \alpha }\right) &= \partial_r \left(\frac{\beta_A}{2 \alpha} \right) \nn \\
\partial_A \left(\frac{\beta_B}{2 \alpha} \right) &= \partial_B \left(\frac{\beta_A}{2 \alpha} \right)~.
\end{align}
It can be verified that these integrability conditions are satisfied courtesy the hypersurface-orthogonality conditions, \ref{HSO1} and \ref{HSO2}, that we have imposed. If we transform to coordinates $\left(t,r, x^A \right)$, we will have
\begin{align}
 \partial_t f\vert_{r,x^A} =\partial_u f \frac{\partial u}{\partial t} \vbar_{r,x^A}=\partial_u f~. 
\end{align}
Hence, the stationarity condition \ref{Killing-2} in coordinates $\left(t,r, x^A \right)$ becomes
\begin{equation}
 \partial_t \alpha =0;\quad \partial_t \beta_{A} =0; \quad \partial_t q_{AB} =0~.
\end{equation}
We are now ready to write down the line element in the static coordinate system. From \ref{dt_exp}, we have 
\begin{equation}
 du = dt+ \frac{dr}{2r\alpha}- \frac{\beta_A dx^A}{2 \alpha}~,
\end{equation}
which when substituted in \ref{GNCline} gives the line element
\begin{equation}
 ds^2= -2 r \alpha dt^2 + \frac{dr^2}{2 r \alpha} - \frac{\beta_A}{\alpha} dr dx^A + \left(q_{AB}+ \frac{r \beta_A \beta_B}{2 \alpha} \right) dx^A dx^B,
\end{equation}
with $\partial_t \alpha =0$, $\partial_t \beta_A=0$ and $\partial_t q_{AB} =0$. This is the GNC line element written in an explicitly static coordinate system.

To transform to the static coordinate system of \cite{Medved:2004ih}, we can first identify $N^2= 2 r\alpha$. The second step would be to install a Gaussian normal coordinate system in the spatial slice by sending out normal geodesics from the $r=0$ surface. (The explicit coordinate transformations to reach this coordinate system, however, is difficult to obtain in closed form.)
\section{Derivations of expressions used in text}\label{Paper04:AppC}

Let us evaluate a couple of expressions we require in \ref{sec:GRtherm}. Noether current for $\xi ^{a}$ has the following expression $J^a (\xi)= \nabla_b \left[J^{ab}\left(\xi\right)\right]$ with $J^{ab}\left(\xi\right)=\nabla^a \xi^b-\nabla^b \xi^a$. We shall make use of the following expression from \cite{Parattu:2013gwb}:
\begin{equation}\label{LieN}
\pounds _{\xi}N^{a}_{ij} = -\nabla_i \nabla_j \xi^a + \frac{1}{2} \left(\delta^a_i \nabla_j\nabla_m \xi^m + \delta^a_j \nabla_i\nabla_m \xi^m\right)- R^a_{jmi}\xi^m~.
\end{equation}
The object $16\pi k_{a}J^{a}(\xi)$ can be evaluated most easily by using the following identity for any two vector fields $u^{a}$ and $v^{a}$:
\begin{align}
16 \pi u_{a}J^{a}(v)-u_{a}g^{ij}\pounds _{v}N^{a}_{ij}=2R_{ab}u^{a}v^{b}=16 \pi v_{a}J^{a}(u)-v_{a}g^{ij}\pounds _{u}N^{a}_{ij}
\end{align}
Applying the above result for the vectors $k^{a}$ and $\xi ^{a}$, we obtain
\begin{align}
2R_{ab}\xi ^{a}k^{b}&=16 \pi k_{a}J^{a}(\xi)-k_{a}g^{ij}\pounds _{\xi}N^{a}_{ij}
\\
&=16 \pi \xi _{a}J^{a}(k)-\xi _{a}g^{ij}\pounds _{k}N^{a}_{ij}
\end{align}
We have chosen the auxiliary vector $k^{a}$ such that $k_{a}=-\nabla _{a}u$. Thus, the Noether current for $k^{a}$ vanishes. Hence, the above equation can be written to yield the value for $16 \pi k_{a}J^{a}(\xi)$ as
\begin{align}
16\pi k_{a}J^{a}(\xi)=k_{a}g^{ij}\pounds _{\xi}N^{a}_{ij}-\xi_{a}g^{ij}\pounds _{k}N^{a}_{ij}
\end{align}
Using \ref{LieN}, the two Lie variation terms can be calculated in a straightforward manner leading to
\begin{align}
k_{a}g^{ij}\pounds _{\xi}N^{a}_{ij}&=-2\pounds _{\xi}N^{u}_{ur}-q^{AB}\pounds _{\xi}N^{u}_{AB}
=-2\partial _{u}\partial _{r}\ln \sqrt{q}+\frac{1}{2}\partial _{u}q_{AB}\partial _{r}q^{AB},
\end{align}
\begin{align}
\xi _{a}g^{ij}\pounds _{k}N^{a}_{ij}=g^{ij}\pounds _{k}N^{r}_{ij}
&=-4\partial _{r}\alpha -2\alpha \partial _{r}\ln \sqrt{q}-2\partial _{u}\partial _{r}\ln \sqrt{q}
+\frac{1}{2}\partial _{u}q_{AB}\partial _{r}q^{AB}
\nonumber
\\
&\phantom{=~}-\beta ^{2}-\frac{1}{\sqrt{q}}\partial _{A}\left(\sqrt{q}\beta ^{A}\right)~.
\end{align}
Using these expressions, we arrive at
\begin{align}
16 \pi k_{a}J^{a}(\xi)&=4\partial _{r}\alpha +2\alpha \partial _{r}\ln \sqrt{q}
+\beta ^{2}+\frac{1}{\sqrt{q}}\partial _{A}\left(\sqrt{q}\beta ^{A}\right)~.
\end{align}
Now, the Noether current expression can be written as
\begin{align}
2R_{ab}\xi ^{a}k^{b}&=16\pi k_{a}J^{a}(\xi)-k_{a}g^{ij}\pounds _{\xi}N^{a}_{ij}
\end{align}
The above relation can also be verified by calculating directly in GNC coordinates:
\begin{align}
-R_{ab}\ell ^{a}k^{b}&=R_{ur}=\partial _{a}\Gamma ^{a}_{ur}-\partial _{u}\Gamma ^{a}_{ar}+\Gamma ^{a}_{ur}\Gamma ^{b}_{ab}-\Gamma ^{a}_{ub}\Gamma ^{b}_{ra}
\nonumber
\\
&=\partial _{r}\Gamma ^{r}_{ur}+\partial _{A}\Gamma ^{A}_{ur}-\partial _{u}\partial _{r}\ln \sqrt{q} +\Gamma ^{a}_{ur}\partial _{a}\ln \sqrt{q}-\Gamma ^{u}_{ub}\Gamma ^{b}_{ru}-\Gamma ^{r}_{ub}\Gamma ^{b}_{rr}-\Gamma ^{A}_{ub}\Gamma ^{b}_{rA}
\nonumber
\\
&=-2\partial _{r}\alpha -\frac{1}{2}\beta ^{2}-\frac{1}{2}\partial _{A}\beta ^{A}-\partial _{u}\partial _{r}\ln \sqrt{q} -\alpha \partial _{r}\ln \sqrt{q}-\frac{1}{2}\beta ^{A}\partial _{A}\ln \sqrt{q}\nn\\ &\phantom{=~}-\Gamma ^{u}_{uA}\Gamma ^{A}_{ru} -\Gamma ^{A}_{ur}\Gamma ^{r}_{rA}-\Gamma ^{A}_{uB}\Gamma ^{B}_{rA}
\nonumber
\\
&=-2\partial _{r}\alpha -\frac{1}{2}\partial _{A}\beta ^{A}-\partial _{u}\partial _{r}\ln \sqrt{q} -\alpha \partial _{r}\ln \sqrt{q}-\frac{1}{2}\beta ^{A}\partial _{A}\ln \sqrt{q}+\frac{1}{4}\partial _{u}q_{AB}\partial _{r}q^{AB}-\frac{1}{2}\beta ^{2}
\end{align}
Expression for $R_{ab}\xi ^{a}k^{b}$ can be rewritten by using the result $R_{ab}\xi ^{a}k^{b}=G_{ab}\xi ^{a}k^{b}-(1/2)R$ and the field equation $G_{ab}=8\pi T_{ab}$ as
\begin{align}
-T_{ab}\xi ^{a}k^{b}&=-\frac{1}{16\pi}\left[16 \pi k_{a}J^{a}(\xi)-k_{a}g^{ij}\pounds _{\xi}N^{a}_{ij}\right]
-\frac{R}{16\pi}
\end{align}
Since energy was defined in terms of Ricci scalar on the 2-surface  for arbitrary static spacetimes \cite{Kothawala:2009kc}, we need a relation between $R$ and $R^{(2)}$. Using which we arrive at
\begin{align}
\left(-T_{ab}\xi ^{a}k^{b}\right)&=\frac{1}{8\pi}\left(R_{ur}-\frac{1}{2}R\right)
\nonumber
\\
&=\frac{1}{8\pi}\Big(-2\partial _{r}\alpha -\frac{1}{2}\partial _{A}\beta ^{A}-\partial _{u}\partial _{r}\ln \sqrt{q} -\alpha \partial _{r}\ln \sqrt{q}-\frac{1}{2}\beta ^{A}\partial _{A}\ln \sqrt{q}+\frac{1}{4}\partial _{u}q_{AB}\partial _{r}q^{AB}
\nonumber
\\
&-\frac{1}{2}R^{(2)}+2\partial _{r}\alpha +2\alpha \partial _{r}\ln \sqrt{q}
+\frac{1}{\sqrt{q}}\partial _{A}\left(\sqrt{q}\beta ^{A}\right)+\frac{3}{4}\beta ^{2}-\frac{1}{2}\beta ^{2}
\nonumber
\\
&-\frac{1}{4}\partial _{u}q_{AB}\partial _{r}q^{AB}-\partial _{u}\ln \sqrt{q}\partial _{r}\ln \sqrt{q}
+\frac{2}{\sqrt{q}}\partial _{u}\partial _{r}\sqrt{q}\Big)
\nonumber
\\
&=\frac{1}{8\pi}\Big(\frac{1}{4}\beta ^{2}-\frac{1}{2}R^{(2)}+\alpha \partial _{r}\ln \sqrt{q}
+\frac{1}{2\sqrt{q}}\partial _{A}\left(\sqrt{q}\beta ^{A}\right)+\frac{1}{\sqrt{q}}\partial _{u}\partial _{r}\sqrt{q}\Big)
\end{align}
Then, integrating over the $u=\textrm{constant},r=\textrm{constant}$ two dimensional surface with integration measure $d^{2}x\sqrt{q}$, we arrive at
\begin{align}
\int d^{2}x \sqrt{q}\left(-T_{ab}\xi ^{a}k^{b}\right)=\int d^{2}x \sqrt{q}\left(\frac{\alpha}{2\pi}\right)
\partial _{r}\left(\frac{\sqrt{q}}{4}\right)-\partial _{r}E
\end{align}
where the energy is defined as
\begin{align}
E\equiv \frac{1}{8\pi}\int d^{2}x dr\sqrt{q}\Big(-\frac{1}{4}\beta ^{2}+\frac{1}{2}R^{(2)}
-\frac{1}{2\sqrt{q}}\partial _{A}\left(\sqrt{q}\beta ^{A}\right)-\frac{1}{\sqrt{q}}\partial _{u}\partial _{r}\sqrt{q}\Big)
\end{align}
which exactly coincides with the result derived in \ref{Paper04:AppA}.



\providecommand{\href}[2]{#2}\begingroup\raggedright\endgroup

\end{document}